\documentclass[conference]{IEEEtran}
\IEEEoverridecommandlockouts
\usepackage{cite}
\usepackage{amsmath,amssymb,amsfonts}
\usepackage{algorithmic}
\usepackage{graphicx}
\usepackage{textcomp}
\usepackage{xcolor}
\usepackage[colorlinks,urlcolor=blue,linkcolor=blue,citecolor=blue]{hyperref}
\usepackage{ragged2e}
\usepackage{tikz}
\usepackage{rotating}
\usepackage{makecell, caption, booktabs}
\usepackage{url}
\usepackage{multirow}

\def\BibTeX{{\rm B\kern-.05em{\sc i\kern-.025em b}\kern-.08em
    T\kern-.1667em\lower.7ex\hbox{E}\kern-.125emX}}

\usepackage{fancyhdr}
\begin{document}

\title{Documenting use cases in the affective computing domain using Unified Modeling Language }

\author{\IEEEauthorblockN{Isabelle Hupont}
\IEEEauthorblockA{\textit{European Commission} \\
\textit{Joint Research Centre (JRC)}\\
Seville, Spain \\
Isabelle.HUPONT-TORRES@ec.europa.eu}
\and
\IEEEauthorblockN{Emilia Gómez}
\IEEEauthorblockA{\textit{European Commission} \\
\textit{Joint Research Centre (JRC)}\\
Seville, Spain \\
Emilia.GOMEZ-GUTIERREZ@ec.europa.eu}
}

\maketitle

\thispagestyle{fancy}

\begin{abstract}

The study of the ethical impact of AI and the design of trustworthy systems needs the analysis of the scenarios where AI systems are used, which is related to the software engineering concept of ``use case" and the ``intended purpose" legal term. However, there is no standard methodology for use case documentation covering the context of use, scope, functional requirements and risks of an AI system. In this work, we propose a novel documentation methodology for AI use cases, with a special focus on the affective computing domain. Our approach builds upon an assessment of use case information needs documented in the research literature and the recently proposed European regulatory framework for AI. From this assessment, we adopt and adapt the Unified Modeling Language (UML), which has been used in the last two decades mostly by software engineers.  Each use case is then  represented by an UML diagram and a structured table, and we provide a set of examples illustrating its application to several affective computing scenarios. 

\end{abstract}


\section{Introduction}

Given the social and ethical impact that some affective computing systems may have~\cite{hupont2022landscape}, it becomes of the utmost importance to clearly identify and document their context of use, envisaged operational scenario or intended purpose.  
Undertaking such use case documentation practices would benefit, among others, system vendors and developers to make key design decisions from early development stages (e.g. target user profile/population, data gathering strategies, human oversight mechanisms to be put in place), authorities and auditors to assess the potential risks and misuses of a system, end users to understand the permitted uses of a commercial system, the people on whom the system is used to know how their data is processed and, in general, the wide public to have a better informed knowledge of the technology. 

The need for transparency and documentation practices in the field of Artificial Intelligence (AI) has been widely acknowledged in the recent literature~\cite{hupont2022documenting}. Several methodologies have been proposed for AI documentation, but their focus is rather on data~\cite{gebru2018datasheets} and models~\cite{mitchell2019model} than AI systems as a whole, limiting at most the documentation of use cases to a brief textual description. Nowadays, voluntary AI documentation practices are in the process of becoming legal requirements in some countries. The European Commission presented in April 2021 its pioneering proposal for the Regulation of Artificial Intelligence, the AI Act~\cite{AIact}, which regulates software systems that are developed with AI techniques such as machine or deep learning. Interestingly, the legal text does not mandate any specific technical solutions or approaches to be adopted; instead, it focuses on the \textit{intended purpose} of an AI system which determines its risk profile and, consequently, a set of legal requirements that must be met. The AI Act's approach further reinforces the need to properly document AI use cases.

The concept of \textit{use case} has been used in classic software development for more than 20 years. Use cases are powerful documentation tools to capture the context of use, scope and functional requirements of a software system. They allow structuring requirements according to user goals~\cite{cockburn2001writing} and provide a means to specify the interaction between a certain software system and its environment~\cite{fantechi2003applications}. 
This work revisits classic software use case documentation methodologies, more particularly those based on the Unified Markup Language (UML) specification~\cite{UML251}, and proposes a template-based approach for AI use case documentation considering current information needs identified in the research literature and the European AI Act. Although  the documentation methodology we propose is horizontal, i.e. it can be applied to different domains (e.g AI for medicine, social media, law enforcement), we address the specific information needs of affective computing use cases.
The objective is to provide a standardised basis for an AI and affective computing technology-agnostic use case repository, where different aspects such as intended users, opportunities or risk levels can be easily assessed. To the best of our knowledge, this is the first methodology specific to the documentation of AI use cases.

The remainder of the paper is as follows. Section~\ref{sec:background} provides an overview of the current AI regulatory framework, existing approaches for the documentation of AI and affective computing systems, and a background on UML. Section~\ref{sec:uml_template} identifies use case information needs and proposes an UML-based methodology for their unified documentation. In Section~\ref{sec:examples}, we put the methodology into practice with some concrete exemplar affective computing use cases. Finally, Section~\ref{sec:conclusions} concludes the paper.

\section{Background}
\label{sec:background}

\subsection{``Intended purpose'' and ``emotion recognition systems'' in the  AI Act}
\label{subsec:aia_intended}

The \textit{intended purpose} of an AI system is central to the European AI Act. It is defined as {\it ``the use for which an AI system is intended by the provider, including the specific context and conditions of use, as specified in the information supplied by the provider in the instructions for use, promotional or sales materials and statements, as well as in the technical documentation''}\footnote{The definitions provided in this manuscript are as of August 2022. The legal text is currently under negotiation and may be subject to change.}. 
An AI system's intended purpose determines its risk profile which can be, from highest to lowest: (1) \textit{unacceptable risk}, covering harmful uses of AI or uses that contradict ethical values; (2) \textit{high-risk}, covering uses identified through a list of high-risk  application areas that may create an adverse impact on people's safety, health or fundamental rights; (3) \textit{transparency risk}, covering uses that are subject to a set of transparency rules (e.g. conversational agents, \textit{deepfakes}); and (4) \textit{minimal risk}, covering all other AI systems. 

The AI Act explicitly and implicitly refers to affective computing systems in several parts of the legal text\footnote{Please note that this assessment is based on the authors' own interpretation of the legal text as of August 2022.}. A transparency risk generally applies to affective computing systems, but there are some clearly identified prohibited practices and high-risk areas. Prohibited practices include systems used to distort a person's behaviour to cause psychological harm, and systems used by public authorities to perform social scoring based on predicted personality or social behaviour. AI systems intended to be used as \textit{``polygraphs and similar tools or to detect the emotional state of a person''} are listed as high-risk in the areas of \textit{``law enforcement''} and \textit{``migration, asylum and border control management''}. There might be situations where emotion recognition is exploited in recruitment contexts or to determine access to educational institutions, which would also be high-risk, as would emotion recognition systems being a safety component of a product (e.g. a system integrated in a car that detects a driver’s drowsiness and undertakes a safety action) or that are part of a machine or medical device (e.g. a companion robot for autistic children).  

Therefore, the AI Act establishes a clear set of harmonised rules that link use cases --including affective computing ones-- to risk levels, which in turn imply different legal requirements. This opens the door to the creation of a use case documentation methodology allowing for an unambiguous assessment of risk levels, such as the one proposed in this work, which could be a valuable tool for different stakeholders, ranging from system providers to authorities.

\subsection{Current approaches for the documentation of AI systems}

In the recent years, both key academic and industry players have proposed methodologies aiming at defining documentation approaches that increase transparency and trust in AI. Among the most successful initiatives, we find some that focus on documenting the datasets used for AI, such as \textit{Datasheets for Datasets}~\cite{gebru2018datasheets}, \textit{The Dataset Nutrition Label}~\cite{holland2018dataset,chmielinski2022dataset} and \textit{Data Cards}~\cite{pushkarna2022data}, as well as some that address the documentation of AI models and algorithms from a  
technical perspective, such as \textit{Model Cards}~\cite{mitchell2019model} and  \textit{AI Factsheets}~\cite{arnold2019factsheets}. Very recently, the Organisation for Economic Co-operation and Development (OECD) has proposed a 
policy-oriented \textit{Framework for the classification of AI systems}~\cite{OECD} to which high-calibre institutions and a large number of AI practitioners have contributed. Being in the form of questionnaires or more visual factsheets, these methodologies are not based on formal documentation standards or specifications. Moreover, even though some of them do explicitly ask about the intended use of AI the system (e.g. \textit{``What is the intended use of the service output?"}\cite{arnold2019factsheets} and \textit{``Intended Use"} section in~\cite{mitchell2019model}), it is just in very broad terms and provided examples lack sufficient details to address complex legal concerns. To date, there is no unified and comprehensive AI documentation approach focusing exclusively on use cases.

\subsection{Documentation of affective computing use cases}

The aforementioned documentation approaches have scarcely been used in the field of affective computing. Only the \textit{Model Cards} original paper comes with a ``smiling detection in images" and a detection of ``toxicity in text" example. Use cases in the field have rather been presented to the community in plain text form (i.e. without following any documentation template), either in survey papers~\cite{aranha2019adapting,weninger2015emotion,zhao2019affective}, in papers presenting a very concrete application~\cite{xu2018automated,murali2021affectivespotlight,setiono2021enhancing} or in articles discussing ethical issues~\cite{hernandez2021guidelines,ong2021ethical,hupont2022landscape}. Interestingly, the Association for the Advancement of Affective Computing (AAAC) has recently launched the \textit{affective computing commercial products database}~\cite{aaac_productdb}, which presents a table with a list of commercial products, a brief description of each one and associated tags such as modality (e.g. speech, text, face), format (e.g. software, hardware) and application domain (e.g. general purpose, education, health). It is however limited to a high level description of real products in the market.

\subsection{Unified Modeling Language (UML) for use case reporting}

\begin{figure}[htb!]
    \centering
    \includegraphics[width=0.65\linewidth]{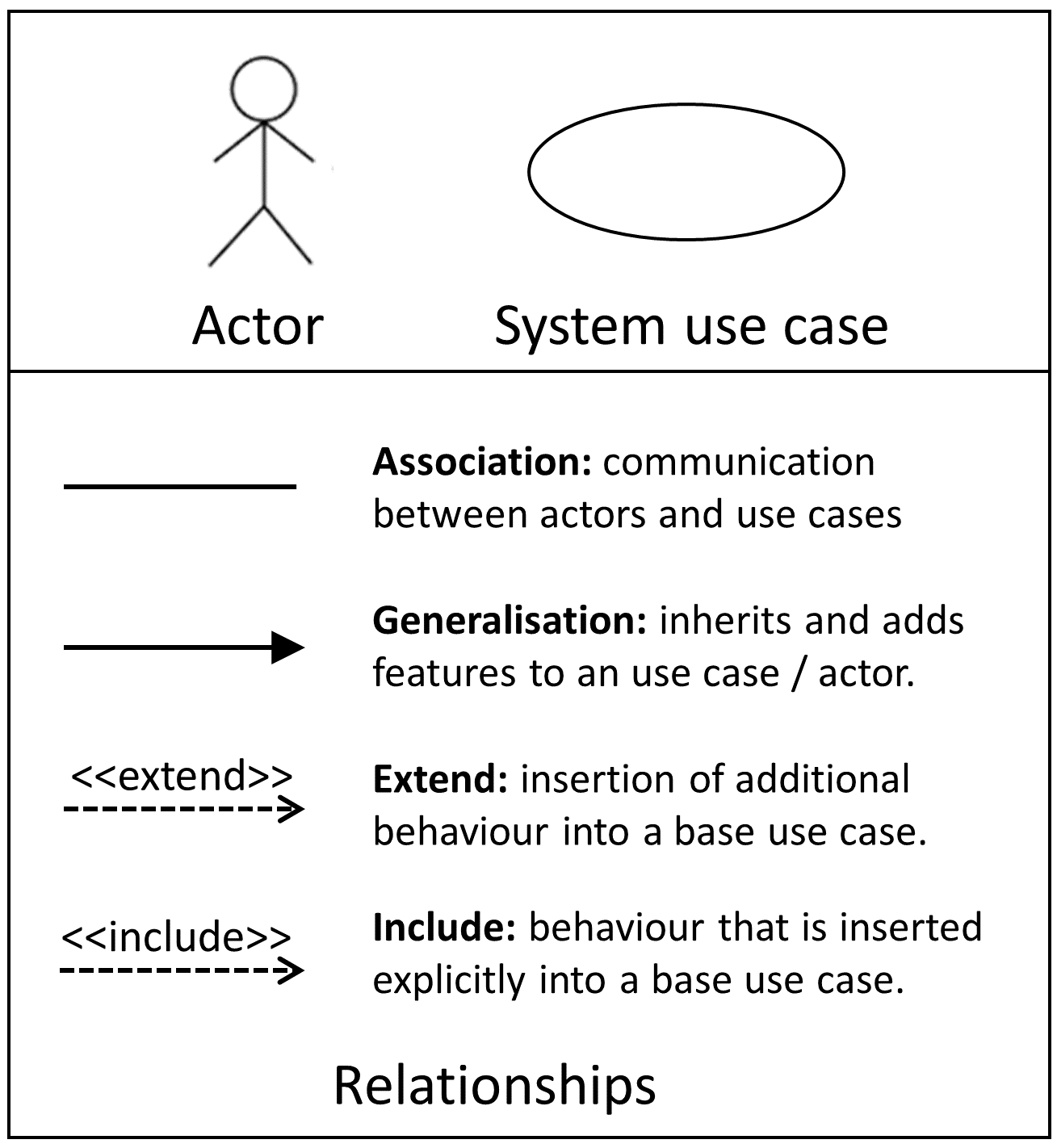}
    \caption{Main UML graphical notation elements for use case modeling.}
    \label{fig:uml_icons}
\end{figure}

The Unified Modeling Language (UML) specification has been widely used in software engineering in the last two decades~\cite{UML251,kocc2021uml}. It provides a standard way to visualize the design and behaviour of a system by introducing a set of graphical notation elements. In particular, it allows for use case modelling, without entering into implementation details, in the form of intuitive \textit{use case diagrams} whose main elements are depicted in Figure~\ref{fig:uml_icons}.

Use cases capture a system's requirements, i.e., what the system is supposed to do. A use case is triggered by an \textit{actor} (it might be a person or group of persons), who is called \textit{primary actor}. The use case describes the various sets of interactions that can occur between the various actors, while the primary actor is in pursuit of a goal. A use case is completed successfully when the goal that is associated with it is reached. Use case descriptions also include possible extensions to this sequence, e.g., alternative sequences that may also satisfy the goal, as well as sequences that may lead to failure in completing the service. 

Once use cases have been modelled in a diagrammatic form, the next step is to describe them in an easy-to-understand and structured written manner. Traditional use case modelling always includes this step, and several standards have been suggested for the layout of use case descriptions. The most widely used is the table format proposed by Cockburn in~\cite{cockburn2001writing} and shown in Figure~\ref{fig:use_case_tables}-left. 

UML is an powerful tool for use case documentation and communication, even to non-technical audiences. Nevertheless, it has not yet been exploited to document AI use cases.

\section{An UML-based documentation methodology for AI and affective computing use cases}
\label{sec:uml_template}

We propose a novel methodology for the documentation of AI use cases which is grounded on (1) the UML standard specification for use case modeling and (2) the requirements for use case documentation under the European AI Act. Our methodology pays particular attention to information needs related to affective computing use cases. It is intended to be a tool to increase transparency and facilitate the understanding of the intended purpose of an AI system, in order to ease the assessment of its risk level and other relevant contextual considerations.

\subsection{Information needs related to the ``intended purpose'' under the AI Act}
\label{sec:aia}

As discussed in Section~\ref{subsec:aia_intended}, the European AI Act centres around the concept of \textit{intended use}. 
Several key information elements are essential to document the intended use of an AI system according to the legal text. We have compiled them in the list presented in Table~\ref{tab:info_elements_aia}. As can be seen, the intended purpose of the system shall be put into context by providing additional information on: who will be the users and the target persons on which the system is intended to be used; the  operational, geographical, behavioural and functional contexts of use that are foreseen, including a description of the hardware on which the system is intended to run (e.g. to highlight whether it is part of a device/machine); which are the system's inputs and outputs; and, if applicable, whether the system is a safety component of a product. Additionally, it is as important to clearly specify the intended use of the system as its foreseeable potential misuses and unintended purposes. 

Finally, the \textit{application areas} information element is one of the most important to assess when it comes to identify a system's risk level. The legal text links some practices, areas and concrete applications within these areas to prohibited practices and high-risk profiles. Table~\ref{tab:areas} compiles the prohibited practices (top) and high-risk application areas (bottom) mentioned in the legal text that are directly related to emotion recognition, or where some kind of affective computing technique could potentially be used (e.g. personality prediction for social scoring, facial expression recognition for student proctoring, pain detection for establishing priority in emergency services). 
In order to facilitate the identification of the level of risk of an affective computing system, it is therefore essential to indicate whether its intended application area(s) or any foreseeable misuse are among those on the list.

It should be noted that Table~\ref{tab:info_elements_aia} is not meant to be a final and exhaustive list of information elements needed for compliance with any future legal requirement. First and foremost, because the AI regulation is still under negotiation, and is therefore subject to be modified in its road towards adoption. Second, because the objective of this work is the documentation of use cases, which is just a small part of the technical documentation required to demonstrate conformity with the legal text.

\newcolumntype{I}{>{\raggedright\arraybackslash}m{0.08\textwidth}}
\newcolumntype{D}{>{\raggedright\arraybackslash}m{0.36\textwidth}}

\begin{table}[h]
\centering
\begin{tabular}{ID}
\textbf{Element} & \textbf{Description}  \\
\toprule 
Intended purpose & Use for which an AI system is intended by the provider. If the system is a safety component of a product, it must be clearly stated.  \\
\midrule
User & Natural or legal person using an AI system under its authority. \\
\midrule
Target persons  & Persons or group of persons on which the system is intended to be used. \\
\midrule
Context of use  & Description of all forms on  which the system is deployed (e.g. characteristics of the specific geographical, behavioural or functional setting) and of the hardware on which it is intended to run.  \\
\midrule
Application areas & List of areas in which the AI system is intended to be applied, including those in Table \ref{tab:areas}.\\
\midrule
Reasonably foreseeable misuses   & Uses of an AI system in a way that is not in accordance with its intended purpose, which may lead to errors, faults, inconsistencies, or risks to health, safety or fundamental rights.    \\
\midrule
Inputs  & Data provided to or directly acquired by the system, on the basis of which the system produces an output. \\
\midrule
Outputs  & Outputs of the AI system as provided to the user.  \\
\bottomrule
\end{tabular}
\caption{Key use case information elements that are needed to assess an AI system's risk level according to the AI Act.}
\label{tab:info_elements_aia}
\end{table}

\newcolumntype{A}{>{\raggedright\arraybackslash}m{0.46\textwidth}}

\begin{table}[h]
\centering
\begin{tabular}{A}
\textbf{AREA  \textcolor{violet}{$>$ POTENTIAL AFFECTIVE COMPUTING USE}}  \\ 
\toprule
- Deploy subliminal techniques beyond a person's consciousness \\
\textcolor{violet}{\hspace{0.75cm} $>$ Distort a person's behaviour to cause psychological harm}\\
- Exploit the vulnerabilities of a specific group of persons \\
\textcolor{violet}{\hspace{0.75cm} $>$ Distort a person's behaviour to cause psychological harm}\\
- Social scoring by public authorities or on their behalf \\
\textcolor{violet}{\hspace{0.75cm} $>$ Evaluation of trustworthiness based on predicted personality}\\ 
\textcolor{violet}{\hspace{0.75cm} $>$ Evaluation of trustworthiness based on social behaviour}\\ 
\toprule
- Education and vocational training \\
\textcolor{violet}{\hspace{0.75cm} $>$ Determine access to educational institutions}\\ 
\textcolor{violet}{\hspace{0.75cm} $>$ Assess students in educational institutions}\\ 
- Employment, workers management and access to self-employment  \\
\textcolor{violet}{\hspace{0.75cm} $>$ Recruitment or selection of natural persons}\\ 
\textcolor{violet}{\hspace{0.75cm} $>$ Make decisions on promotion/termination of contract}\\
\textcolor{violet}{\hspace{0.75cm} $>$ Monitoring and evaluation of performance and behaviour}\\
- Access to essential private/public services and benefits \\
\textcolor{violet}{\hspace{0.75cm} $>$ Evaluate eligibility of natural persons for public assistance}\\ 
\textcolor{violet}{\hspace{0.75cm} $>$ Evaluate creditworthiness of natural persons}\\
\textcolor{violet}{\hspace{0.75cm} $>$ Establish priority in the dispatching of emergency services}\\
- Law enforcement   \\
\textcolor{violet}{\hspace{0.75cm} $>$ Make individual risk assessments of natural persons}\\ 
\textcolor{violet}{\hspace{0.75cm} $>$ Detect the emotional state of a natural person}\\
\textcolor{violet}{\hspace{0.75cm} $>$ Crime profiling of natural persons}\\
- Migration, asylum and border control management  \\
\textcolor{violet}{\hspace{0.75cm} $>$ Make individual risk assessments of natural persons}\\ 
\textcolor{violet}{\hspace{0.75cm} $>$ Detect the emotional state of a natural person}\\
\textcolor{violet}{\hspace{0.75cm} $>$ Examine applications for asylum/visa/residence}\\
- Administration of justice and democratic processes   \\
\textcolor{violet}{\hspace{0.75cm} $>$ Assist judicial authority in researching and interpreting facts}\\
\bottomrule
\end{tabular}
\caption{Practices and application areas listed as \textit{prohibited} (top) and \textit{high-risk} (bottom) in the AI Act, that are directly or that could be indirectly related to affective computing. Please note that this table has been generated by the authors based on their own interpretation of the AI Act as of August 2022. }
\label{tab:areas}
\end{table}

\subsection{Revisiting UML for AI and affective computing use case documentation}
\label{sec:revisit_uml}

The idea of \textit{intended use} defined in the  AI Act is closely related to the traditional software concept of \textit{use case}, as defined in the UML specification. UML use case diagrams do not enter into technical details (e.g. implementation details, algorithm architectures) but rather focus on the context of use, the main actors using the system, and actor-actor and actor-system interactions, which is a focus  aligned with that proposed by the AI Act to assess a system's risk level. The UML language is thus a powerful, standardized and highly visual tool to operationalise the need for a unified documentation of AI use cases. 

\begin{figure*}[htb!]
    \centering
    \includegraphics[width=\linewidth]{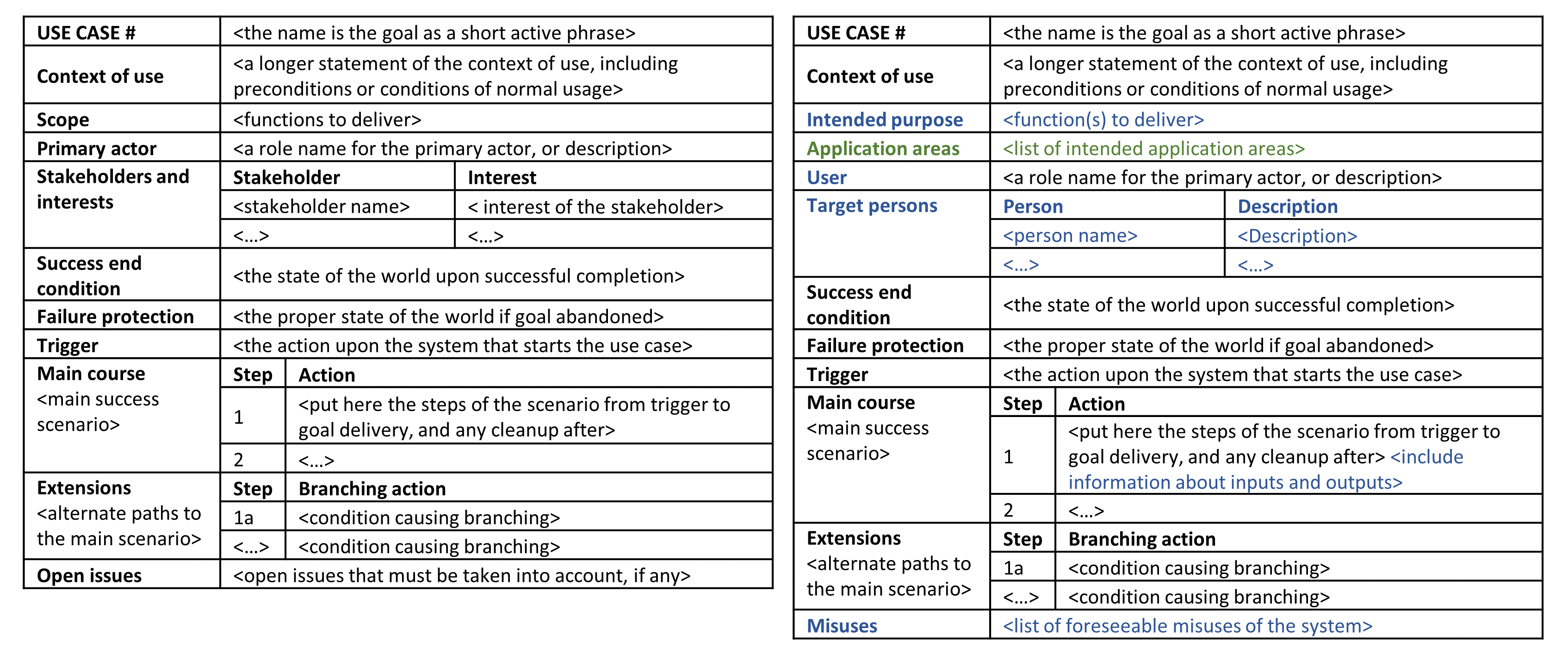}
    \caption{Left: classic table template for the documentation of UML use case diagrams, as in~\cite{cockburn2001writing}. Right: proposed adaptation for the documentation of AI use cases, inspired by the European AI Act's definitions. Green text corresponds to added fields, while blue text is used for fields that have been adapted.}
    \label{fig:use_case_tables}
\end{figure*}

In Figure~\ref{fig:use_case_tables}-right, we propose an adaptation of the classic table template accompanying UML use case diagrams~\cite{cockburn2001writing} to the AI Act's taxonomy. As can be seen, the adapted fields are minimal and there is an almost perfect correspondence with the original template. We have only renamed some key words (in blue in the table), namely \textit{scope} to \textit{intended purpose}, \textit{primary actor} to \textit{user}, \textit{stakeholders and interests} to \textit{target persons}, and \textit{open issues} to \textit{misuses}. We have also included a new field called \textit{application areas} (in green), allowing to clearly identify the area(s) in which the system is intended to be used and, if applicable, specify whether they correspond to those listed in Table~\ref{tab:areas}.

\section{Methodology in practice: example of affective computing use cases}
\label{sec:examples}

\begin{figure*}[h!]
    \centering
    \includegraphics[width=\linewidth]{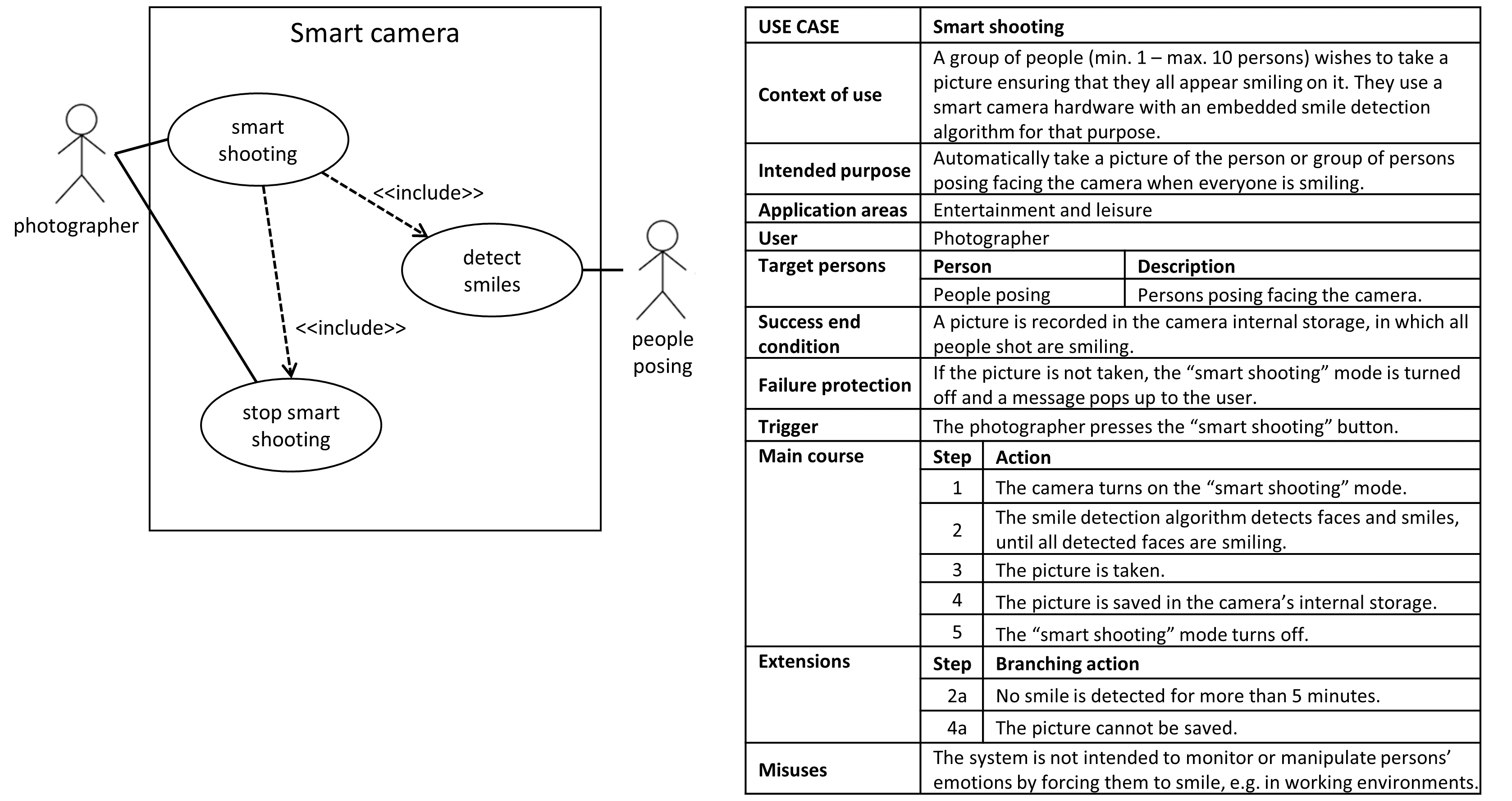}
    \caption{First use case: methodology applied to a smart camera system with embedded smile detection capabilities.}
    \label{fig:uc_smile}
\end{figure*}

\begin{figure*}[h!]
    \centering
    \includegraphics[width=\linewidth]{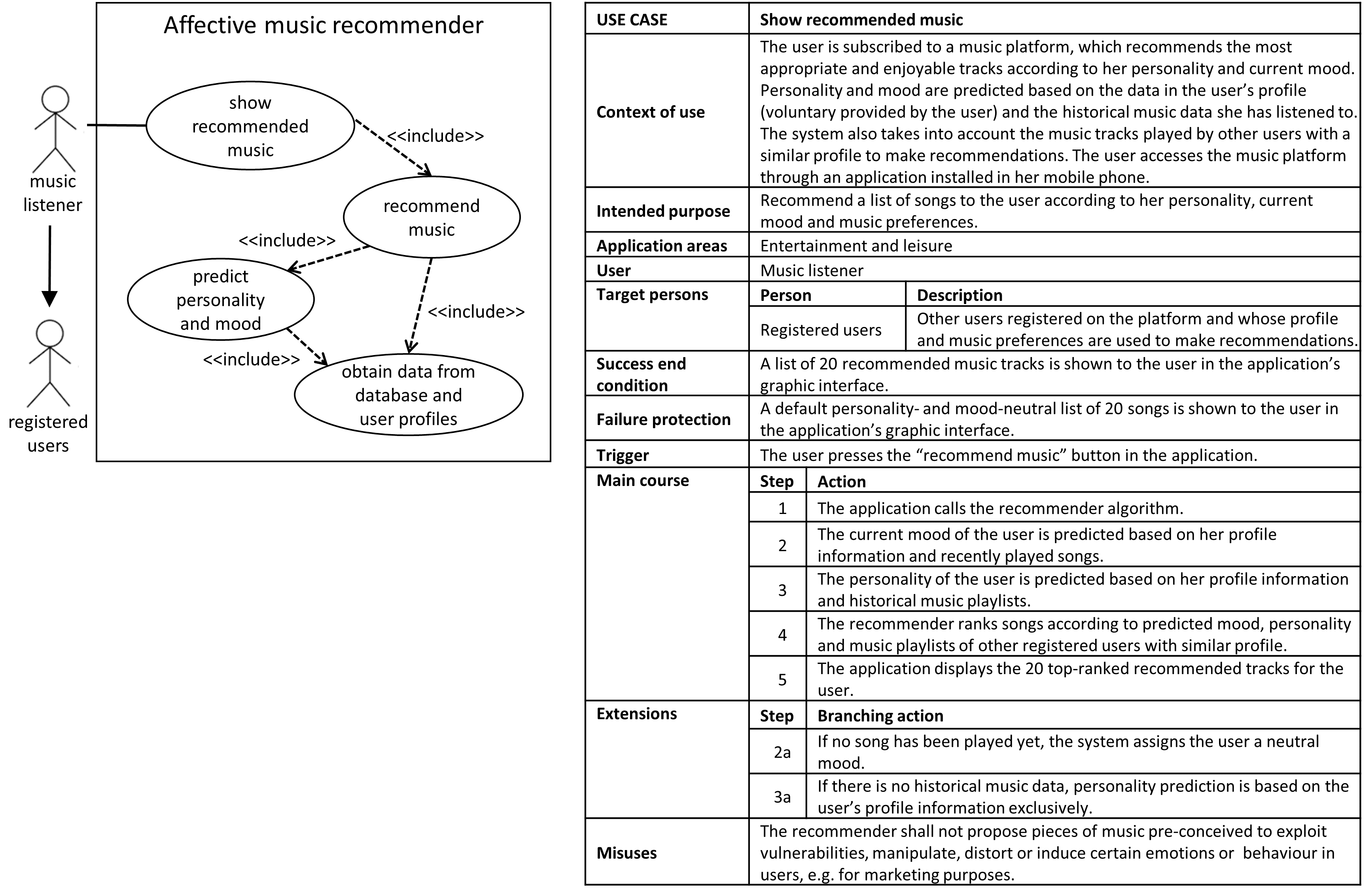}
    \caption{Second use case: proposed methodology applied to an affective music recommender system.}
    \label{fig:uc_music}
\end{figure*}

\begin{figure*}[h!]
    \centering
    \includegraphics[width=\linewidth]{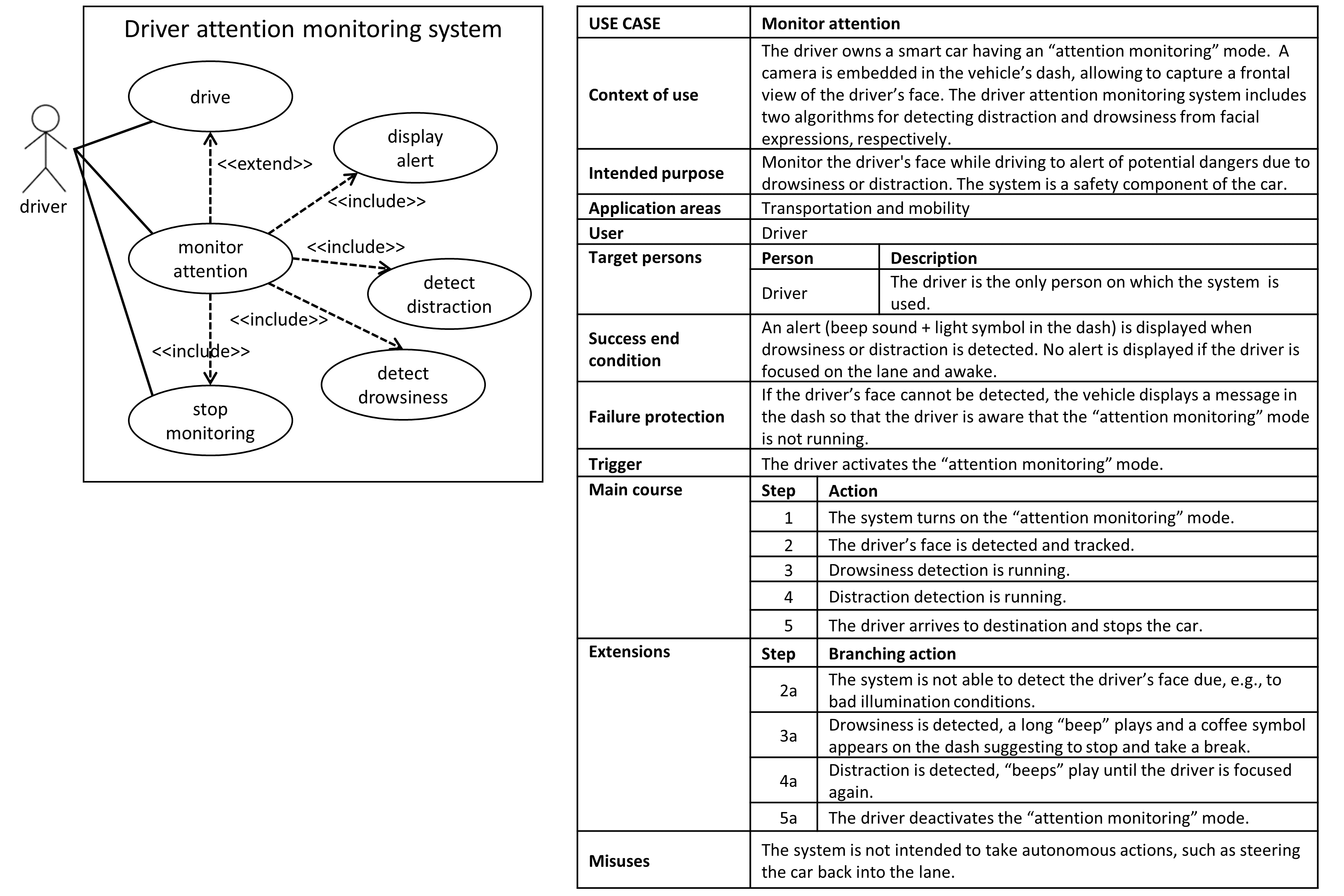}
    \caption{Third use case: proposed methodology applied to a driver attention monitoring system.}
    \label{fig:uc_car}
\end{figure*}

In this section, we apply the proposed methodology to the documentation of three representative affective computing systems. Figures~\ref{fig:uc_smile}-\ref{fig:uc_car} show their corresponding UML use case diagrams and accompanying tables, which are further described below.   \\

\noindent \textbf{Smart camera}. In this first use case, the system is a smart camera that shoots a picture only when all the people posing in front of it are smiling. There are several products in the market with this feature~\cite{canon,nikon}, which have inspired this example. The UML diagram of the \textit{smart shooting} use case and its corresponding table are shown in Figure~\ref{fig:uc_smile} left and right, correspondingly. This application may seem simple and naive a priori, but it has recently caused controversy. Workers at a Beijing office were forced to smile to an AI camera to get through the front doors, change the temperature or print documents, in an attempt to improve the working environment by keeping workers happy~\cite{news_happyface}. However, some workers felt their emotions were manipulated. Our proposed UML table makes it clear that the target application domain is \textit{entertainment and leisure} exclusively, and the \textit{misuses} field explicitly emphasises that the system is not conceived to be used to monitor or manipulate emotions in contexts such as working environments. This important claim excludes the use case from the high-risk area of \textit{workers management $>$ monitoring and evaluation of performance and behaviour} (c.f. Table~\ref{tab:areas}).  \\

\noindent \textbf{Affective music recommender}. Figure~\ref{fig:uc_music} shows the UML diagram and table for the second use case, corresponding to an affective music recommender system proposing songs to the user based on her personality, current mood and playlist history. This use case has been inspired by the work presented in~\cite{amini2019affective}. Several studies have shown that users' music playlists can be used to infer  emotions, personality traits and vulnerabilities~\cite{deshmukh2018survey}; the other way round, certain music pieces can induce behaviours and manipulate listeners' emotions~\cite{gomez2021music}. The proposed methodology allows to frame the ethical use of the system by documenting step by step its conceived functioning, and how and for what purpose personality and mood prediction are extracted and used (based on profile data voluntarily provided by the platform's users, with the sole purpose of making the most appropriate and enjoyable music recommendations). The \textit{misuses} field further strengthens the system's ethical principles by explicitly signaling the prohibition of proposing music pre-conceived to exploit vulnerabilities, manipulate, distort or induce certain emotions or behaviour in users, which would be a \textit{prohibited practice} according to the AI Act (c.f. Table~\ref{tab:areas}).\\

\noindent \textbf{Driver attention monitoring}. The third example is a use case where a driver's face is recorded with a car in-cabin camera, and monitored in order to recognise drowsiness and distraction. When such situations are detected, the vehicle's attention monitoring system sends alerts in the form of beep tones and light symbols in the car dash (Figure~\ref{fig:uc_car}). Driver monitoring systems have been a popular affective computing application in the last decade  and the modelling of this use case is inspired by different papers~\cite{kumar2018driver,govindarajan2018affective} as well as real commercial products~\cite{subaru,tesla}. The \textit{intended purpose} field in the proposed UML-based table clearly states that the system is part of a safety component of the vehicle, which immediately positions it as a high-risk profile according to the AI Act. Further, the documentation methodology allows to indicate that the system is conceived to alert the driver, but in any case to allow the vehicle to take full control of the car in an autonomous manner. 

\section{Conclusions and future work}
\label{sec:conclusions}

In this paper, we propose a methodology for the documentation of AI use cases which covers the particular information elements needed to  address affective computing ones. The methodology has a solid grounding, being based on two strong pillars: (1) the UML use case modelling standard, and (2) the recently proposed European AI regulatory framework. Each use case is represented in a highly visual way by means of an UML diagram, accompanied by a structured and concise table that compiles  the relevant information to understand the intended use of a system, and to assess its risk level and foreseeable misuses. Our approach is not intended to be an exhaustive methodology for the technical documentation of AI or affective computing systems (e.g. to demonstrate compliance with legal acts). Rather, it aims to provide a template for compiling related use cases with a simple but effective and unified language, understandable even by non-technical audiences. We have demonstrated the power of this language through practical affective computing exemplar use cases. 

In the near future, we plan to develop a collaborative repository compiling a catalogue of AI --including affective computing-- use cases following the proposed template. The first step will be to transcribe the 60 facial processing applications presented in~\cite{hupont2022landscape}, which contain 18 emotion recognition use cases, in order to add them to this catalogue.

\section*{Ethical Impact Statement}


The methodology presented in this paper proposes the first unified documentation approach for AI use cases, with a strong focus on affective computing ones, which allows to differentiate intended uses and potential misuses. In the last years, the  need for trustworthy AI has been raised by both private and public key institutions and researchers in the field~\cite{OECD,gebru2018datasheets,arnold2019factsheets,madaio2020co,hupont2022landscape}. In particular, documentation has been identified as a key factor towards the fulfilment of \textit{transparency}~\cite{hupont2022documenting}, one of the seven pillar requirements for trustworthy AI established by the  High-Level Expert Group on Artificial Intelligence (AI HLEG)~\cite{HLEG}. Therefore, this work represents a major step towards ethical AI and affective computing, and could even constitute a basis for the future standardisation activities in this area.

\section*{Acknowledgment}

This work is partially supported by the European Commission under the HUMAINT project of the Joint Research Centre.

\bibliographystyle{IEEEtran}
\bibliography{IEEEabrv,biblio.bib}

\end{document}